\documentclass[12pt,a4paper]{article}
\usepackage[latin1]{inputenc}
\usepackage{amsmath}
\usepackage{amsfonts}
\usepackage{amssymb}
\usepackage{graphicx}
\author{Tehseen Rug}
\title{Holography in Cascading DGP}
 \renewcommand\appendix{\par
   \setcounter{section}{0}% 
   \setcounter{subsection}{0}% 
   \setcounter{figure}{0}%
   \renewcommand\thesection{\Alph{section}}% 
   \renewcommand\thefigure{\Alph{section}.\arabic{figure}}}
\begin{document}
\begin{center}
\LARGE{\textbf{Holography of Species in Cascading DGP}}
\end{center}
\begin{center}
\large{\textbf{Tehseen Rug}$^{a,b,1}$}
\end{center}
\vspace*{1cm}
\begin{center}
\textit{$^a$Arnold Sommerfeld Center for Theoretical Physics\\
Department f\"ur Physik, Ludwig-Maximilians-Universit\"at M\"unchen Theresienstr. 37, 80333 M\"unchen, Germany}
\end{center}
\begin{center}
\textit{$^b$Max-Planck-Institut f\"ur Physik\\
F\"ohringer Ring 6, 80805 M\"unchen, Germany}
\end{center}
\vspace*{1cm}
\begin{center}
\textbf{Abstract}
\end{center}
In this paper we generalize the idea of "species holography" to the case of Cascading DGP theories. Essence of the phenomenon is that a 4D field theory with $N$ particle species coupled to a high-dimensional bulk gravity propagating solely a graviton becomes strongly coupled at the scale that would be the quantum gravity scale of the bulk theory if all the species were propagating in the bulk. We will see that both, crossover scales and Vainshtein scales, can be understood as holographic scales in the above sense. We confirm our results by an explicit effective field theoretic derivation of these scales in Cascading DGP.
\vspace*{\fill}

$^1$Tehseen.Rug@physik.uni-muenchen.de
\section{Introduction}
The idea that certain properties of field theories can be understood in terms of higher-dimensional  gravity theories and vice-versa is fascinating and intriguing.\\
\hspace*{0.5cm}
The origin of this holographic principle goes back to some considerations of 't Hooft concerning the nature of black holes \cite{t'Hooft}. According to the holographic principle, it should be possible to describe black hole physics in $D+1$ spacetime dimensions using the holographic degrees of freedom defined on the boundary of the black hole.\\
\hspace*{0.5cm}
The most well-known example of such a holographic connection between higher-dimensional and lower-dimensional theories is the AdS/CFT correspondence \cite{Maldacena, Witten}.\\ 
The question, whether it is possible to extend the idea of AdS/CFT and to find new dualities between gravity and field theory, is of fundamental importance both from a conceptual and computational point of view.\\
\hspace*{0.5cm}
In this paper we shall discuss another intriguing example of holographic connection which we shall refer to as "species holography" or "strong coupling holography". This idea was originally suggested in \cite{Holography}. The essence of the phenomenon is the following: The strong coupling scale of a theory on a brane with $N$ propagating degrees of freedom coupled to bulk gravity is given by the scale that would be a higher-dimensional gravity cut-off if all the species were propagating in the bulk. This connection holds even in an appropriate gravity decoupling limit. Thus, effective field theory on the brane knows about its gravitational origin through the strong coupling scale of the theory. In \cite{Holography} this effect was studied in the context of the DGP model \cite{DGP1}, theories with solitonic domain walls and string theory with D-branes.\\ 
\hspace*{0.5cm}
We will generalize the arguments given in \cite{Holography} for the DGP model to cascading gravity \cite{Cascade1, Cascade2}. We will find that the strong coupling scales of the field theory are still given by the cut-off scales of higher-dimensional gravity theory with $N$ holographic degrees of freedom. Further, the crossover scales of  Cascading DGP can also be expressed in terms of the number of particle species. This suggests a deep, fundamental principle that has yet to be understood.\\
\hspace*{0.5cm}
The paper is organized as follows: In the next section, we will review how the cut-off scale \cite{BlackHoles1}
\begin{align}
M_{N(4+d)} \equiv \frac{M_{4+d}}{N^{\frac{1}{2+d}}}
\end{align}
arises in a theory of gravity and $N$ different particle species in $4+d$ dimensions. Here, $M_{4+d}$ is the Planck mass in $4+d$ dimensions. In particular, the implications of this cut-off for braneworld scenarios will be discussed.\\
In the third section, we will repeat the basic ideas of the Cascading DGP model using effective field theory methods.\\
In the fourth section we will combine our findings of the first and second section. Namely, we will show, how various scales that appear in the field-theoretic treatment of the Cascading DGP model in six dimensions can be understood as holographic scales. The most important result is that the phenomenon of "strong coupling holography" persists. We will then generalize our findings to Cascading DGP in arbitrary dimensions.\\
In the last section, the results will be summarized and discussed.
\section{The Black Hole Bound}
In this section we will explain, how the bound (1) arises relying on the consistency of the black hole physics in $4+d$ dimensions in the presence of $N$ particle species. Here we will focus on the information storage derivation of the bound \cite{Information}\footnote{The same bound was also found in the recently formulated Black Hole Quantum N Portrait \cite{NPortrait}. This picture gives a microscopic understanding of the black hole as a Bose condensate of weakly interacting gravitons. Thus, the bound (1) can also be derived from microscopic quantum considerations.}.\\
In order to establish the cut-off (1), one can perform the following thought experiment: Suppose you want to localize all the $N$ particle species in a box of size $L_{box}$. The species being different means that the quantum numbers of an individual species distinguishes it from the other particles. In order to obtain information about all the $N$ species in the box, we need to require that the size of the box is larger than its gravitational radius. Below its Schwarzschild radius, the box collapses into a black hole and the resolution of the individual species is no longer possible.\\
Clearly, the localization of an individual species on a lengthscale $L_{box}$ costs at least the energy $L_{box}^{-1}$. Thus, a box which contains the wavefunctions of all the $N$ species has mass $M_{box} = N/L_{box}$. The corresponding Schwarzschild radius in $4+d$ dimensions is then given by 
\begin{align}
r_{box} = \Bigl( \frac{M_{box}}{M_{4+d}^{d+2}} \Bigr)^{\frac{1}{d+1}} = \Bigl( \frac{N}{L_{box}}\frac{1}{M_{4+d}^{d+2}} \Bigr)^{\frac{1}{d+1}}.
\end{align}
Demanding that the size of the box is bigger than its Schwarzschild radius fixes the lower bound
\begin{align}
L_{box} \geq L_{N(4+d)}.
\end{align}
This implies that it is not possible to localize all the species within a lengthscale smaller than the scale $L_{N(4+d)} \equiv M_{N(4+d)}^{-1}$. In the other words, below this distance the particles cannot be treated as elementary anymore and some change of regime in the field theory must occur.\\
It follows that in a theory with $N$ species, the fundamental lengthscale is no longer the corresponding Planck scale, but rather the scale $L_{N(4+d)}$. We shall call this scale the holographic species scale, due to the fact that the scaling of $N$ coincides with the area of a $4+d$-dimensional black hole measured in units of the Planck area. Thus, the length  $L_{N(4+d)}$ saturates the lowest bound on the information-storage for $N$-bits, and thus resonates with the 't Hoofts holographic approach to black hole physics.  This connection has deeper roots since the species label is a particular way of quantum-information storage.\\
\hspace*{0.5cm}
We now wish to investigate how the picture changes when the species are localized on a lower-dimensional subspace. In the simplest case this analysis was done in \cite{Holography} and the conclusion was that the lower-dimensional field theory inevitably exhibits a strong coupling phenomenon exactly at the holographic scale of higher-dimensional gravity even in the decoupling limit in which the theory becomes purely lower-dimensional and non-gravitational.\\
\hspace*{0.5cm}
We will now review the original argument which suggests a connection between different scales appearing in the problem. Later we will apply our considerations to a more subtle case which is known as Cascading DGP.\\\\
Suppose you have a theory where $N$ particle species are localized on a k-brane, i.e. a $k+1$-dimensional hypersurface which is embedded in a $D$-dimensional bulk spacetime, $k+1 < D$. The brane is taken to be static and infinitely thin. Further, suppose that only gravity can propagate in the directions transverse to the brane which are taken to be infinitely extended.\\
This implies that the cut-off of the bulk gravity theory is given by the $D$-dimensional Planck length $L_D$. Since the localized particles interact with the higher dimensional graviton, the cut-off on the brane is given by $L_{ND}$.\\
This leads to the following conclusion:\\
Either the species theory breaks down at lengthscales smaller than $L_{ND}$, or, from the point of view of an observer living on the brane, gravity changes at a scale $r_c > L_{ND}$ and becomes weaker. In this case, the fundamental scale for the brane observer would be lowered to the corresponding higher dimensional Planck scale.\\
Let us elaborate the two possible modifications of the theory.\\
Breakdown of the species theory means that it is not possible to store information about all the species on a scale smaller than $L_{ND}$. Below this scale, the species cannot be treated as elementary anymore, but, for example, the species could become composite states.\\
If gravity seems to become weaker for a brane observer, it will be possible to localize information about all the species on smaller scales than the cut-off $L_{ND}$. To be more precise, change of gravity means, that from the point of view of an observer on the brane, the gravitational force law and the associated gravitational radius will shift away from $D$-dimensional behaviour. This would then resolve the puzzle that the two observers experience different cut-offs. In particular, in such a scenario, gravity compromises in such a way that while an asymptotic higher-dimensional observer sees $D$-dimensional gravity all the way till $L_D$, the observer on the brane sees gravity that crosses over from $D$ to $k+1$-dimensional behavior. A well-known example for such a crossover behaviour of gravity is the DGP model with $k=3$ and $D=5$. The striking feature in this model is that the strong-coupling scale, which is characteristic for helicity-zero graviton is given by a higher-dimensional holographic scale. The argument will be repeated and generalized to Cascading DGP later on. We will see that the phenomenon of strong coupling holography persists in the Cascading DGP model.\\
\hspace*{0.5cm}
In the Cascading DGP model, as we will see in the next section, the second possibility, i.e. the modification of gravity at some crossover scales will be realized naturally. We will show how these scales are fixed by the requirement, that the cut-off of the higher-dimensional gravity theory and the cut-off for the species theory on the brane should be the same in section 4. Further, we will see a connection between $L_{ND}$ and strong coupling on the brane.\\
Before starting this discussion, we will discuss some important properties of the Cascading DGP model in the next section. 
\section{Cascading DGP}
The Cascading DGP model is a consistent higher-dimensional generalization of the original DGP model \cite{Cascade1, Cascade2}. In order to understand the basic properties of the model, we will restrict the discussion to the 6D cascading set-up at first. In particular, in the first part of this section, we will review the most important properties of the model. In the second part of this section, we will give an explicit field theoretic derivation of the second Vainshtein scale that appears in Cascading DGP.\\
\hspace*{0.5cm}
The idea of the model is that there is a 3-brane on which all the $N$ matter particles are localized. This brane is embedded in a 4-brane which lives in a 6D bulk spacetime which is asymptotically flat. Both branes are taken to be static and infinitely flat. The gravitational part of the action is given by
\begin{align}
S_{cas} &= M_6^4 \int d^4 x d y d z \sqrt{-g_{(6)}} R_6 + M_5^3 \int d^4 x d y \sqrt{-g_{(5)}} R_5 \nonumber \\ &+ M_4^2 \int d^4 x \sqrt{-g_{(4)}} R_4.
\end{align}
The $M_n, g_n, R_n; n= 4, 5, 6$ are the 4D, 5D and 6D Planck masses, metrics and Ricci scalars, respectively. Both extra dimensions are infinitely extended and only gravity has access to these extra dimensions. Computing the scalar part of the propagator, one can see that gravity is modified at very large distances. The propagator has an interpolating behaviour, i.e. it gives rise to 4D gravity at small distances, then 5D gravity at intermediate scales, and finally 6D gravity at the largest distances provided that the two crossover scales $\rho_c = M_5^3/M_6^4$ and $r_c = M_4^2/M_5^3$ fulfil $r_c < \rho_c$. We will assume that this condition holds in what follows. In the case $r_c > \rho_c$ a direct crossover from 6D to 4D gravity takes place. This case will be discussed briefly in section 4.\\
For consistency with observations, both crossover scales have to be taken to be bigger than the current day Hubble horizon.\\
\hspace*{0.5cm}
Just as any other theory of infrared modification of gravity, the Cascading DGP model is subject to the famous van-Dam Veltman Zakharov (vDVZ) discontinuity of the graviton propagator at the linearized level \cite{vDVZ}. The essence of the discontinuity is that the appearance of extra degrees of freedom due to extra dimensions (or due to a massive 4D graviton) leads to a discrepency between the graviton propagator of IR modified gravity theories and general relativity. Even in a suitable limit (e.g. sending the mass of the graviton to zero in massive gravity or sending the crossover scales in Cascading DGP to infinity) one cannot recover the propagator derived from general relativity. This seems worrysome, but as shown in \cite{Strong Interactions2} all IR-modfied theories of gravity exhibit strong coupling. This suggests that the perturbative expansion breaks down at very large distances. The scale at which non-linearities become important is the so-called Vainshtein scale $r_*$. Because of strong coupling, the full, non-linear solution reproduces GR in a certain limit.\\
\hspace*{0.5cm}
In the Cascading DGP model, one has to deal with the tensor structure of 6D gravity. This implies that the consistency of the model requires two Vainshtein effects, the first one turning 6D into 5D gravity and the second one turning 5D into 4D gravity.
The Vainshtein scales can be found most easily using the method of boundary effective action. This was done first in \cite{Strong Interactions1} for the original DGP model. The idea is to integrate out the bulk and to find an effective action valid on the 3-brane. Keeping only the leading non-linearities, it is possible to isolate the strong coupling phenomenon which is responsible for the appearance of the Vainshtein effect.
In the DGP model the Vainshtein radius is found to be 
\begin{align}
r_* = (r_c^2 r_g)^{1/3}
\end{align}
where $r_g$ is the gravitational radius of a given spherically symmetric source and $r_c = M_4^2/M_5^3$ is the crossover scale.\\
In what follows, we will adopt this method for the Cascading DGP model following \cite{Cascade1, Cascade2}. Starting from the action (4) one can integrate out the bulk. Keeping only the leading non-linearities leads to an effective action valid on the 4-brane:
\begin{align}
\mathcal{L}^{5D}_{dec} = \frac{M_5^3}{4} h^{AB}(\mathcal{E} h)_{AB} - 3 M_5^3 \partial^A \pi \partial_A \pi + \frac{27 M_5^3}{32 \mu_c^2} \partial^A \pi \partial_A \pi \square_5 \pi.
\end{align}
The first term is simply the linearized Einstein-Hilbert Lagrangian in 5D with the linearized Einstein tensor 
\begin{align}
\mathcal{E}_{CD}^{AB} h_{AB} &= \square_5 h_{CD} - \square_5 \eta_{CD} h_A^A - \partial^A \partial_D h_{AC} \nonumber
\\ &- \partial^A \partial_C 
h_{AD} + \eta_{CD} \partial^A \partial^B h_{AB} + \partial_C \partial_D h_A^A.
\end{align}
$A, B \ldots$ denote five-dimensional indices, $\mu_c \equiv \rho_c^{-1}$ and $\square_5$ is the d'Alembertian in five dimensions. The second term is the kinetic energy of the scalar part of the graviton,
\begin{align}
\pi \eta_{AB} = \tilde{h}_{AB} - h_{AB},
\end{align}
with $\tilde{h}_{AB}$ the full physical metric fluctuation. The important term is the interaction term in (6). After canonically normalizing the scalar field, we can read off a strong coupling scale from the interaction term,
\begin{align}
\Lambda_{strong} = (M_6^{16} M_5^{-9})^{\frac{1}{7}}.
\end{align}
Higher non-linearities vanish in the gravity decoupling limit, where one sends $M_6, M_5 \rightarrow \infty$ keeping the strong coupling scale fixed. Thus, the strong coupling effect is fully captured in the leading cubic interaction of the scalars. In order to find the Vainshtein scale connected to the 6D-to-5D crossover one can derive the equations of motions for $\pi$ assuming a coupling of this field to the trace of the energy-momentum tensor $T_{\mu \nu}, \; \mu, \nu = 0, 1, 2, 3$ localized on the codimension 2 brane. One finds \cite{Cascade2} that the interacting piece becomes as important as the kinetic term exactly at the scale
\begin{align}
\rho_* = \Bigl( \frac{M}{\mu_c^2 m_c M_4^2}  \Bigr)^{\frac{1}{4}},
\end{align}
where $\mu_c \equiv M_6^4/M_5^3$, $m_c \equiv M_5^3/M_4^2$ and $M$ is the mass of the source. This sets the Vainshtein scale for the 6D-to-5D transition. Below that scale, the scalar interactions decouple from the gravitational spin-2 interactions on the 4-brane. This implies, that for the distances smaller than $\rho_c$, the model essentially reduces to the original DGP model. Therefore one would expect that the second Vainshtein scale is the same as in the DGP model,
\begin{align}
r_* = (r_c^2 r_g)^{\frac{1}{3}}. \nonumber
\end{align}
We shall now explicitly derive this second Vainshtein radius. In order to do so, one needs to find an effective 4D theory containing kinetic terms and the leading non-linearities. The derivation of this effective theory already exists in the literature \cite{Bigalileon}. The idea is to non-linearly complete the Lagrangian (6) and to integrate out the fifth dimension keeping the leading non-linearities. The result for the scalar part is (the spin-2 contribution is the usual linearized Einstein-Hilbert Lagrangian in four dimensions)
\begin{align}
\mathcal{L} &= \frac{1}{2} \chi \square_4 \chi + \alpha \pi \square_4 \pi - \frac{1}{3 \sqrt{6} \mu_{strong}^3} \partial_{\mu} (\chi + \pi) \partial^{\mu} (\chi + \pi) \square_4 (\chi + \pi) \nonumber \\ &+ \frac{1}{\sqrt{6}} (\chi + \pi) \frac{T}{M_4}.
\end{align}
Here $\chi + \pi$ is the scalar, canonically normalized part of the physical graviton, $\alpha$ is an order one, positive numerical constant depending on the regularization scheme\footnote{In a naive treatment, there is a ghost in the cascading set-up. This ghost, however, can be cured in many different ways, see e.g. \cite{Cascade1, Cascade2}.}, $T$ is the trace of the energy-momentum tensor on the brane and $\mu_{strong} \equiv (m_c^2 M_4)^{1/3}$ is the strong coupling scale.\\
It can be shown that all higher-order non-linearities vanish in the gravity decoupling limit,
\begin{align}
M_4, M_5 \rightarrow \infty
\end{align}
with $\mu_{strong}$ fixed.\\
Starting from (11) we will now derive the Vainshtein scale, i.e. the scale at which the non-linear term in (11) becomes as important as the kinetic terms. This is equivalent to the scale at which the long-distance and short-distance behaviour of the potential created by the scalars cross. Thus, we need to solve the equation of motion for the first derivatives of the fields.
Upon variation of (11) one finds the following equations of motion:
\begin{align}
\alpha &\square_4 \pi - c [(\partial_{\mu} \partial_{\nu}(\pi + \chi))^2 - (\square(\pi + \chi))^2] = - \frac{T}{\sqrt{6} M_4} \nonumber \\ &\square_4 \chi - c [(\partial_{\mu} \partial_{\nu}(\pi + \chi))^2 - (\square(\pi + \chi))^2] = - \frac{T}{\sqrt{6} M_4}
\end{align}
Here we defined $c \equiv \frac{2}{3 \sqrt{6} \mu_{strong}^3}$.\\
These equations can be rewritten into a total divergence,
\begin{align}
&\partial^{\mu}\Bigl[\alpha \partial_{\mu} \pi + c \partial_{\mu}(\pi + \chi) \square_4 (\pi + \chi) - \frac{c}{2} \partial_{\mu}(\partial(\pi + \chi))^2 \Bigr] = - \frac{T}{\sqrt{6} M_4} \nonumber \\ &\partial^{\mu} \Bigl[\partial_{\mu} \chi + c \partial_{\mu}(\pi + \chi) \square_4 (\pi + \chi) - \frac{c}{2} \partial_{\mu}(\partial(\pi + \chi))^2 \Bigr] = - \frac{T}{\sqrt{6} M_4}.
\end{align}
Taking a point-like source, $T = -M \delta^3(x)$, (14) can be integrated. The result is
\begin{align}
\alpha &M_4 \pi' + 2 c \frac{M_4}{r} (\pi' + \chi')^2 = \frac{M_4^2 r_g}{\sqrt{6} r^2} \nonumber \\ &M_4 \chi' + 2 c \frac{M_4}{r} (\pi' + \chi')^2 = \frac{M_4^2 r_g}{\sqrt{6} r^2},
\end{align}
where $'$ denotes the derivative with respect to $r$.\\
In terms of the physical (not canonically normalized) scalar metric fields \cite{Bigalileon} $\chi, \pi \rightarrow \sqrt{2/3}\frac{\chi, \pi}{M_4}$ the solution for the first derivatives is then found to be 
\begin{align}
\pi' + \chi' = \sqrt{\frac{2}{3}} \frac{1}{M_4} \frac{\alpha^2 r}{4 c (\alpha + 1)}\Biggl(-1 + \sqrt{1 + \frac{8 c M_4 r_g (\alpha +1)^2}{\sqrt{6} \alpha^2 r^3}} \Biggr).
\end{align}
From this result one can see that the potential for the scalars behaves as $r^{-1/2}$ for small distances. This implies that the gravitational effect produced by the scalars in suppressed compared to the Einsteinian spin-2 contribution in the vicinity of astrophysical sources. For large distances, however, the solution scales as $r^{-2}$ implying that the scalar contribution becomes important at large scales. The Vainshtein radius is then determined to be the scale at which the two behaviours cross. Up to unimportant order one numerical constants, one finds
\begin{align}
r_* = (r_c^2 r_g)^{\frac{1}{3}}, \nonumber
\end{align}
in agreement with our expectations.\\\\
The fact, that there are two Vainshtein scales suggests that the Cascading DGP model actually reduces to general relativity at small distances. This makes the model a reasonable candidate for infrared modification of gravity.\\
\hspace*{0.5cm}
Having reviewed the basic ideas of the effective field theory treatment briefly, we will turn our attention to holography in the next section. As we will see, it will be possible to derive the hierarchy of Planck scales in terms of the number of particle species $N$ using non-perturbative arguments. It will also be possible to confirm the Vainshtein scales, (5) and (10), using strong coupling holography.
\section{Holography in Cascading DGP}
We will now combine the knowledge of the previous two sections to obtain information about Cascading DGP in terms of black hole physics. As we will see, we will be able to establish the hierarchy of the different Planck scales in terms of the number of particle species localized on the 3-brane. Further, it will be discovered that the holographic correspondence between the higher-dimensional holographic cut-off and the Vainshtein scale found for the original DGP model \cite{Holography} will persist in the Cascading DGP model. The fact, that a similar relation was also found for string theoretic D-branes and field theories with domain walls \cite{Holography} suggests that there should be a deep underlying principle which is yet to be understood\footnote{In fact, strong coupling holography within the framework of string theory and D-branes follows from the AdS/CFT correspondence, where the role of $N$ is played by the number of colours \cite{Holography}.}.\\
\hspace*{0.5cm}
At first, we will focus on the case $\rho_c > r_c$.
As we saw in section 2, having the same cut-off on the field theory side on the 3-brane with $N$ particle species and in the bulk with only gravity suggests that gravity, from the point of view of an observer living on the brane, must change. In the 6D set-up the crossover must set in at a scale $\rho_c > L_{N6}$. In the Cascading DGP model, we know, that gravity changes from 6D to 5D at this crossover scale. This implies that for the distances $r_c < r < \rho_c$ an observer on the 3-brane will experience the laws of 5D gravity. Having the same cut-off then yields
\begin{equation}
L_{N5} \equiv N^{\frac{1}{3}} L_5 = L_6.  
\end{equation}
From this equation, we can see how the hierarchy between the Planck lengths $L_5$ and $L_6$ in terms of the number of particle species is generated. The crossover scale is then determined to be
\begin{equation}
\rho_c = \frac{M_5^3}{M_6^4} = N L_{6}. 
\end{equation}
This tells us that indeed, $\rho_c > L_{N6} = N^{1/4} L_6$, as required. We can now go a step further. After the first crossover, one basically has a situation similar to the original DGP case. Insisting that the cut-off of the field theory on the brane should hold even up to smaller distances than $L_6$ suggests that another crossover must take place. This crossover must occur at a distance $r_c > L_{N5}$ We know that this is exactly what happens in the Cascading DGP model. Demanding that the cut-off on the 3-brane and the cut-off on the 4-brane which is given by $L_5$ since only gravity can propagate on that brane should coincide, leads to 
\begin{equation}
L_{N4} \equiv \sqrt{N} L_4 = L_5. 
\end{equation}
This fixes the second crossover scale to be 
\begin{equation}
r_c = \frac{M_4^2}{M_5^3} = N L_{5}
\end{equation}
which indeed satisfies the condition $r_c > L_{N5}$.\\
Combining equations (17) and (19) tells us that one has the following hierarchy of Planck scales,
\begin{equation}
M_4 \gg M_5 \gg M_6.
\end{equation}
In order to generate a reasonable hierarchy between the fundamental Planck scale, $M_6$, and the electroweak scale, $M_{EW}$, we choose $M_6 \sim 1$ TeV. Having $M_4 \sim 10^{19}$ GeV, we find that the number of particle species on the 3-brane should be given by $N \sim 10^{24}$. Using (19), 5D Planck mass is then given by $M_5 \sim 10^{11}$ GeV.\\
The huge number of species needed to generate a sensible hierarchy could be realized using string theoretic constructions. At low energies such constructions often involve a large number of particle species.\\
\hspace*{0.5cm}
Having discussed the hierarchy problem, we will now try to understand in what sense there is a connection between the field theory on the brane and higher-dimensional gravity. Following \cite{Holography} where it was shown that the Vainshtein scale for a single bit of information in 5D in the original DGP model is given by the higher-dimensional gravity cut-off for $N$ bits of information, $r_* = L_{N5}$, we will now assume that this strong coupling holography should also hold in the Cascading DGP model. As we will see, this will lead to the same Vainshtein scales that were discussed in section (3).\\
\hspace*{0.5cm}
Before going to the Cascading DGP model, we will briefly review why the relation $r_* = L_{N5}$ holds in DGP. For that purpose, consider a single bit of information in 5D of size $L_5$ with mass $M_5$. This is the smallest possible semi-classical black hole in the bulk, since only gravity has access to the extra dimension in DGP. From the 5D perspective, the size of this black hole is the same as its Schwarzschild radius $r_{g5}$. Since the Vainshtein effect takes place on the brane at distances smaller than $r_c$, i.e. at lengthscales were gravity effectively looks four-dimensional on the brane, we need to determine the 4D gravitational radius of this single bit of information. Using equation (2) for the case $M_{box} = M_5$ and $d = 0$ and equation (20), we find
\begin{equation}
r_{g4} = \frac{r_{g5}}{N}.
\end{equation}
The corresponding Vainshtein scale for such a source is then given by
\begin{equation}
r_* = (r_c^2 r_{g4})^{\frac{1}{3}} = N^{\frac{1}{3}} L_5 \equiv L_{N5}.
\end{equation}
This tells us that the Vainshtein scale for a single bit of information in the DGP model is set by the bulk holographic species scale for $N$ bits of information.\\
\hspace*{0.5cm}
With this preparation, we can move on and apply similar arguments to the Cascading DGP model.
The basic assumption is that the Vainshtein scale corresponding to the 6D-to-5D crossover, $\rho_*$, should be given by the bulk holographic scale in six dimensions for $N$ bits of information, $L_{N6}$. Indeed, we find
\begin{align}
\rho_* = N^{\frac{1}{4}} L_6 = (N M_6^3 M_6^{-7})^{\frac{1}{4}} = \Bigl( \frac{M_6}{\mu_c^2 m_c M_4^2}  \Bigr)^{\frac{1}{4}}.
\end{align}
This is the same scale as (10) for a single bit of information in six dimensions. The analysis presented here, however, is fully non-perturbative and independent of any dynamics. It is rather only based on the consistency of the black hole physics.\\
The second Vainshtein scale can then be found just as in the DGP case. It is again determined by $L_{N5}$.\\
\hspace*{0.5cm}
To summarize this analysis, we saw that strong coupling holography persists in the Cascading DGP model. This suggests that even in the gravity decoupling limit which is performed in the field theoretic treatment, the higher-dimensional gravitational origin is not lost in the effective description. The memory of higher-dimensional gravity is somehow encoded in the strong coupling phenomenon in the field theoretic approach. As already mentioned, a similar holography also holds for the solitonic domain walls and string theoretic D-branes. We think that this cannot be an amazing coincidence. Rather, there should be some underlying principle relating the field theory and gravity description. In particular, this suggests the deep underlying connection between the large distance modification of gravity in four dimensions and a higher-dimensional 
gravity in terms of holography of species. To put it shortly, the large distance modification of gravity can be viewed as a necessity of the theory to accommodate the difference in the number of higher-dimensional and lower-dimensional species in consistency with black hole physics. In particular, the origin of the strong coupling in large distance modification of gravity can be understood in terms of holography of species.\\
\hspace*{0.5cm}
We will now discuss the case $r_c > \rho_c$. In this case gravity directly changes from six-dimensional to four-dimensional behaviour at a scale $r_c^{dir} = (\mu_c m_c)^{- \frac{1}{2}}$ \cite{Cascade2}. Keeping this in mind, we see that the scales $\rho_c$ and $r_c$ can no longer be interpreted as crossover scales.\\
In terms of the species number a direct transition implies $r_c^{dir} = N^{\frac{1}{2}} L_6$\footnote{This directly implies that the transition sets in before the holographic cut-off $L_{N6}$ is reached. This is in accordance with our argumentation.}. Thus,
\begin{align} 
N^{\frac{1}{2}} L_6 = (\mu_c m_c)^{- \frac{1}{2}} = \frac{M_4}{M_6^2}.
\end{align}
As expected from the direct transition, we immediately find a natural hierarchy between the 4D and 6D Planck masses:
\begin{align} 
M_4 = N^{\frac{1}{2}} M_6.
\end{align}
To establish a complete hierarchy of Planck scales in this scenario, we need to use the fact that the 6D bulk holographic scale is bigger than that of the 5D theory, $L_{N6} > L_{N5}$ and $\rho_c < r_c.$  The first condition yields
\begin{align}
M_5 > N^{\frac{1}{12}} M_6.
\end{align}
The  second condition gives $M_5^6 < M_6^4 M_4^2$. This is readily rewritten into
\begin{align}
M_5 < N^{-\frac{1}{3}} M_4.
\end{align}
Having $N \gg 1$ particle species localized on the 3-brane, the inequalities (27) and (28) display the expected hierarchy between the Planck masses:
\begin{align}
M_4 \gg M_5 \gg M_6.
\end{align}
This is again a similar hierarchy as was found before in the case $r_c < \rho_c$. It is however not possible to express the hierarchy in terms of strict equalities as in the case discussed above.\\
As a next step, we can look for strong coupling within this scenario. The direct crossover from 6D to 4D gravity implies the existence of only one such scale in contrast to the scenario discussed before. Below this scale the tensor structure of the linearized graviton propagator should mimic the usual Einsteinian one, while above that scale the tensor structure corresponds to 6D gravity. Relying on holography we now insist that the Vainshtein scale $r_*^{dir}$ is set by the holographic bound of the bulk gravity theory, i.e.
\begin{align}
r_*^{dir} \equiv L_{N6}.
\end{align}
Using $r_{g6} = L_6$ and the definition of $r_c^{dir}$ one easily finds
\begin{align}
r_*^{dir} = \Bigl( (r_c^{dir})^2 (r_{g6})^2 \Bigr)^{\frac{1}{4}}.
 \end{align}
This scale can be interpreted as the scale at which the interactions of two scalars become as important as their kinetic term in the effective field theory description on the 3-brane, thereby decoupling from the gravitational interactions. Of course, the results derived here should be confirmed using field theoretic methods. Such an analysis, however, is out of the scope of this paper.\\\\
\hspace*{0.5cm}
We will now generalize the results to Cascading DGP in arbitrary dimensions. The idea is to embed branes with increasing dimensionality into each other. Thus, a 3-brane sits in a 4-brane and so on until finally a $4+(d-3)$-brane is placed into a $4+(d-2)$-brane. This geometric set-up then lives in a $4+d$-dimensional bulk spacetime. Branes of all dimensionalities, as well as the bulk, get their own Einstein-Hilbert terms. This set-up leads to $d$ different crossover scales, $r_{4+k}$, $k = 1,2 \ldots d$, at which gravity changes dimensionality by 1 at each crossover, provided that the scales are ordered in the following way:
\begin{align}
r_{4+k} > r_{4+(k-1)}.
\end{align}
We will assume that such an ordering holds. Otherwise there could be direct crossovers as discussed before for the 6D Cascading DGP model. We will not consider such direct crossover in what follows.\\
The crossovers must take place before the bulk holographic cut-off is reached, i.e.
\begin{align}
r_{4+k} > L_{N(4+k)}, \; \; \; k=1, 2, \ldots d \; \; \; L_{N(4+k)} = N^{\frac{1}{2+k}} L_{k+4}.
\end{align}
Here $L_{k+4} \equiv 1/M_{4+k}$ is the $4+k$-dimensional Planck length.\\
Since we know that at each crossover scale the behaviour of the gravitational force law changes by one power of $r$ we can express the scales also in terms of the number $N$ of particle species:
\begin{align}
r_{4+k} = NL_{4+k}.
\end{align}
Using this relation and
\begin{align}
r_{4+k} = \frac{M_{4+(k-1)}^{2+(k-1)}}{M_{4+k}^{2+k}},
\end{align}
we can relate the different Planck masses to each other\footnote{Note that these crossover scales can be found by analyzing the scalar part of the propagator. For $k = 1, 2$, these are simply the scales $r_c$ and $\rho_c$.},
\begin{align}
M^{2 +(k-1)}_{4 +(k-1)} = N M^{2+(k-1)}_{4+k}.
\end{align}
We can now derive a general expression for the Vainshtein scales. Following the discussions of the cases $d=1, 2$ discussed before, we propose that at each stage the Vainshtein scale is given by the corresponding holographic cut-off, i.e.
\begin{align}
r^k_* \equiv L_{N(4+k)}.
\end{align}
This can be rewritten in terms of Planck scales,
\begin{align}
r^k_* = (M^{2 + (k-1)}_{4 +(k-1)} M^{- 2 k- 3}_{4+k})^{\frac{1}{2+k}}.
\end{align}
For $d=1, 2$, equation (38) gives the same Vainshtein scales found before in the original DGP model and in the 6D cascading set-up.\\
\hspace*{0.5cm}
This discussion completes our holographic picture of the Cascading DGP model.
\section{Discussion}
The purpose of this paper was to provide one more connection between the consistencies of seemingly unrelated phenomena: strong coupling effects in 4D field theory and holographic scales dictated by consistency of black hole physics in higher-dimensional gravity.\\
\hspace*{0.5cm}
We have demonstrated that the required hierarchy of Planck scales in Cascading DGP can be understood in terms of consistency of lower-dimensional theories with large number of species with black hole physics. The crossover scales $\rho_c$ and $r_c$ could be fixed by demanding that the observers on the brane and in the bulk should experience the same cut-off.\\
We then extended the analysis of strong coupling holography \cite{Holography} to Cascading DGP. It was possible to confirm the expectation that strong coupling holography holds in this scenario as well. In particular, we showed that the Vainshtein scales $\rho_*$ and $r_*$ which appear in the field theoretic treatment are given by the bulk holographic species scales for $N$ bits of information, $L_{N6}$ and $L_{N5}$. Combining our results with the ones given in \cite{Holography}, we expect some fundamental principle which relates field theory on a brane to gravity in the bulk, provided that the bulk geometry is asymptotically flat. Studying the dynamics of localization of species on the brane could be one way to get a deeper insight into strong coupling holography.\\
\hspace*{0.5cm}
Assuming that strong coupling holography holds, we also extended our analysis to Cascading DGP in arbitrary dimensions. The fact that there is a Vainshtein scale for each holographic scale (provided that the crossover scales are ordered as in (32)), suggests that all higher-dimensional generalizations reduce to general relativity at small distances, since each Vainshtein scale changes the tensor structure of the linearized graviton propagator. To be more precise, one starts with the tensor structure of $4+d$-dimensional gravity at the largest distances. The first Vainshtein scales changes the structure to $4+(d-1)$ gravity. This goes on until, finally, the last Vainshtein scale turns 5D into 4D gravity.\\
\hspace*{0.5cm}
To conclude, strong coupling holography appears to be a surprisingly powerful tool which can be used to obtain information about strong dynamics on the brane.  According to our findings field theories on the brane localized in asymptotically flat high-dimensional spaces exhibit connection of strong coupling scales between brane and bulk theory which can be understood in terms of number of species that the two theories propagate. The connection is established through black hole physics, but persists even in the gravity decoupling limit. This suggests that the strong coupling scales of a class of non-gravitational theories may have a hidden gravitational interpretation in terms of invisible embedding in higher-dimensional gravity theory.
\section{Acknowledgements}
It is a pleasure to thank Gia Dvali for valuable discussions about species holography. Further, I would like to thank Humboldt Foundation and the International Max Planck Research School on Elementary Particle Physics for support. The work of T.R. was supported by Humboldt Foundation.


\newpage
\begin{thebibliography}{10cm}
\bibitem[1]{t'Hooft} G. 't Hooft, {\em "Dimensional Reduction in Quantum Gravity"}, (1993), arXiv:gr-qc/9310026v2
\bibitem[2]{Maldacena} J.M. Maldacena, {\em "The Large N Limit of Superconformal Field Theories and Supergravity"}, Adv.Theor.Math.Phys.2:231-252 (1998); arXiv:hep-th/9711200
\bibitem[3]{Witten} E. Witten, {\em "Anti De Sitter space and Holography"}, Adv.Theor.Math.Phys.2:253-291 (1998); arXiv:hep-th/9802150
\bibitem[4]{Holography} G. Dvali, C. Gomez, {\em "Strong Coupling Holography"}, (2009), arXiv:0907.3237v1
\bibitem[5]{DGP1} G. Dvali, G. Gabadadze, M. Porrati, {\em "4D Gravity on a Brane in 5D Minkowski Space"}, Phys.Lett.B485:208-214 (2000); arXiv:hep-th/0005016v2\\\\
G. Dvali, G. Gabadadze, M. Kolanovic, F. Nitti {\em "Scales of Gravity"}, Phys.Rev.D65:024031 (2002); arXiv:hep-th/0106058v2\\\\
G. Dvali, G. Gabadadze, {\em "Gravity on a Brane in Infinite-Volume Extra Space"}, Phys.Rev. D63 065007 (2001); arXiv:hep-th/0008054v3
\bibitem[6]{Cascade1} C. de Rham, G. Dvali, S. Hofmann, J. Khoury, O. Pujolas, Michael Redi, A. J. Tholley, {\em "Cascading DGP"}, Phys.Rev.Lett.100:251603 (2007); arXiv:0711.2072v2
\bibitem[7]{Cascade2} C. de Rham, S. Hofmann, J. Khoury, A. J. Tholley, {\em "Cascading Gravity and Degravitation"}, JCAP 0802:011 (2008); arXiv:hep-th/0712.2821v1
\bibitem[8]{BlackHoles1} G. Dvali, {\em "Black Holes and Large N Species Solution to the Hierarchy Problem"}, (2007), arXiv:0706.2050\\\\
G. Dvali, M. Redi, {\em "Black Holes Bound on the Number of Species and Quantum Gravity at LHC"}, Phys.Rev.D77:045027 (2008); arXiv:0710.4344
\bibitem[9]{Information} G. Dvali, C. Gomez, {\em "Quantum Information and Gravity Cutoff in Theories with Species"}, Phys.Lett.B674:303-307 (2009); arXiv:0812.1940
\bibitem[10]{NPortrait} G. Dvali, C. Gomez, {\em "Black Hole's Quantum N-Portrait"}, (2011), arXiv:1112.3359v1
\bibitem[11]{vDVZ} H. van Dam, M. Veltman, {\em "Massive and massless Yang-Mills and gravitational fields"}, Nucl. Phys. \textbf{B22}, 397 (1970)\\\\
V.I. Zakharov, {\em "Linearized Gravitation Theory and the Graviton Mass"}, JETP Lett. \textbf{12}, 312 (1970)
\bibitem[12]{Strong Interactions2} G. Dvali, {\em "Predictive Power of Strong Coupling in Theories with Large Distance Modified Gravity"}, NewJ.Phys.8:326 (2006); arXiv:hep-th/0610013v1 
\bibitem[13]{Strong Interactions1} M. A. Luty, M. Porrati, R. Rattazzi, {\em "Strong Interactions and Stability in the DGP Model"}, JHEP 0309:029 (2003); arXiv:hep-th/0303116v1
\bibitem[14]{Bigalileon} A. Padilla, P. M. Saffin, S-Y. Zhou, {\em "Bi-galileon theory I: motivation and formulation"}, JHEP 1012:031 (2010); arXiv:1007.5424v4
\end{thebibliography}
\end{document}